\documentclass[twocolumn,showpacs,amsfonts,aps,prl,floatfix,nofootinbib,float,superscriptaddress]{revtex4-1} 

\usepackage{graphicx}  
\usepackage{dcolumn}   
\usepackage{bm}        
\usepackage{amssymb}   
\usepackage{amsmath}   
\usepackage{tabularx}
\usepackage{placeins}
\usepackage{mathtools}
\usepackage{color}

\def\eq#1{(\ref{#1})}
\def\Eq#1{Eq.~(\ref{#1})}

\newcommand{\beq}{\begin{equation}}
\newcommand{\eeq}{\end{equation}}
\newcommand{\bea}{\begin{align}}
\newcommand{\eea}{\end{align}}
\newcommand{\beas}{\begin{align*}}
\newcommand{\eeas}{\end{align*}}

\newcommand{\Tint}[1]{{\hbox{$\sum$}\!\!\!\!\!\!\!\int\,}_{\!\!\!\!\raise-0.9ex\hbox{$\scriptstyle{#1}$}}}

\begin{document}
\widetext

\title{Thermal dynamics on the lattice with exponentially improved
  accuracy} \author{J.~M.~Pawlowski\,} 
\affiliation{Institute for Theoretical Physics, Universit\"at
  Heidelberg, Philosophenweg 12, D-69120 Germany}
\affiliation{ExtreMe Matter Institute EMMI, GSI, Planckstr. 1,
  D-64291 Darmstadt, Germany}
\author{A.~Rothkopf\,}
\affiliation{Institute for Theoretical Physics, Universit\"at
  Heidelberg, Philosophenweg 12, D-69120 Germany}

\begin{abstract}
  We present a novel simulation prescription for thermal quantum
  fields on a lattice that operates directly in imaginary frequency
  space. By distinguishing initial conditions from quantum dynamics it
  provides access to correlation functions also outside of the
  conventional Matsubara frequencies $\omega_n=2\pi n T$.  In
  particular it resolves their frequency dependence between $\omega=0$
  and $\omega_1=2\pi T$, where the thermal physics $\omega\sim T$ of
  e.g.~transport phenomena is dominantly encoded. Real-time spectral
  functions are related to these correlators via an integral transform
  with rational kernel, so their unfolding is exponentially improved
  compared to Euclidean simulations. 

  We demonstrate this improvement within a $0+1$-dimensional scalar
  field theory and show that spectral features inaccessible in
  standard Euclidean simulations are quantitatively captured.
\end{abstract}

\maketitle

Modern experiments ranging from heavy-ion collisions at RHIC and LHC
\cite{Muller:2013dea,2016EPJP..131...52A} to ultracold quantum gases
\cite{Bloch:RevModPhys.80.885} explore the physics of
strongly-correlated quantum systems across vastly separated
temperature scales $(T\sim 10^{12}..10^{-9}K)$
\cite{2012NJPh...14k5009A}. A unifying aspect of these studies are
thermal phenomena, such as the transport of conserved charges
\cite{Heinz:2011kt,Ryu:2015vwa,Meyer:2007dy,Kamikado:2013sia,Haas:2013hpa,Christiansen:2014ypa,Aarts:2014nba,Mages:2015rea,Ding:2016hua,Chien1504.02907,Cao:2010wa,Enss:2010qh}
or the in-medium modification of heavy bound states
\cite{Ding:2012sp,Aarts:2014cda,Kim:2014iga,Borsanyi:2014vka,Burnier:2015tda,PhysRevLett.100.053201,Nishida:2008ra}.
Connecting these to fundamental interactions, such as QED and QCD,
requires a thorough theoretical understanding of the equilibrium
properties of quantum fields. Unfortunately in realistic experimental
environments many approximate techniques, such as perturbation theory,
are inapplicable and a non-perturbative numerical computation is
required.

Conventional simulations of quantum fields, e.g.\ lattice QCD, are
based on non-perturbative Monte-Carlo techniques relying on an
analytic continuation of the real-time axis into the complex plane
\cite{Foulkes:2001zz,Smit:2002ug}, for real-time approaches to spectral functions see 
e.g.~\cite{Aarts:2001qa,Strauss:2012dg,Kamikado:2013sia,Haas:2013hpa,Tripolt:2013jra,Haas:2013hpa,Christiansen:2014ypa,Helmboldt:2014iya,Pawlowski:2015mia,Yokota:2016tip,Kamikado:2016chk,Jung:2016yxl,Ilgenfritz:2017kkp}.

Two conceptual challenges plague the necessary reconstruction of real-time 
correlation functions from those simulated in imaginary time: the crucial one
is the finite extent of the Euclidean time axis with $\tau\in
[0\,,\,1/T]$. It leads to discrete Matsubara frequencies
$\omega_n=2\pi n T$, which limits the resolution of imaginary
frequencies. In turn, increasing the number of temporal points
$N_\tau$ simply increases the maximum frequency available while
keeping the spacing fixed. Thus, a large $N_\tau$ does not improve the
access to the relevant regime $\omega\sim T$, where thermal
physics is dominantly encoded in $G(\omega)$. The signal to noise
ratio for thermal contributions actually decays rapidly above
$\omega\approx T$ and indeed, already $G(\omega_1=2\pi T)$ may only be
marginally relevant for thermal physics.

In order to resolve this conceptual issue, we present a simulation
algorithm, which operates in imaginary frequency space with arbitrary
resolution, given by the number of frequency points $N_\omega$. In
particular this approach resolves frequencies between $\omega_0$ and
$\omega_1$. 

The second issue concerns the unfolding of spectral functions
$\rho(\mu)$ from imaginary time correlators $G(\tau)$, or imaginary
frequency correlators $G(\omega)$. In Euclidean time the inverse
problem reads\newline\vspace{-0.4cm}
\begin{equation}
  G(\tau)=\int d\mu\, \tilde{K}(\mu,\tau,T)\rho(\mu), \quad
  \tilde{K}(\mu,\tau,T) \overset{\mu\tau >> 1}{\sim} e^{-\mu\tau}\,. 
\label{eq:inverseexp}\end{equation}
\Eq{eq:inverseexp} leads to an exponentially hard ill-posed problem
\cite{JARRELL1996133}. In this work we simulate directly in imaginary
frequency space, where numerical data and spectral functions are
related via the rational K\"all\'en-Lehmann kernel,
\begin{equation}
G(\omega)=\int d\mu\, K(\mu,\omega)\rho(\mu), \quad |K(\mu,\omega)| 
\overset{\mu/\omega >> 1}{\sim} \frac{1}{\mu/\omega}\,.\label{Eq:FreqRec}
\end{equation}
Its polynomial decay significantly improves the success of the
spectral reconstruction already for Mastubara correlators with
$\omega=\omega_n$. Such a rational kernel is moreover an analytic
implementation of the idea behind the Backus-Gilbert
\cite{Brandt:2015sxa} or Sumudu \cite{Pederiva:2014qea} approaches.

\begin{figure*}[t]
\centering
\includegraphics[scale=0.59, trim= 0 0.0cm 0 0, clip=true]{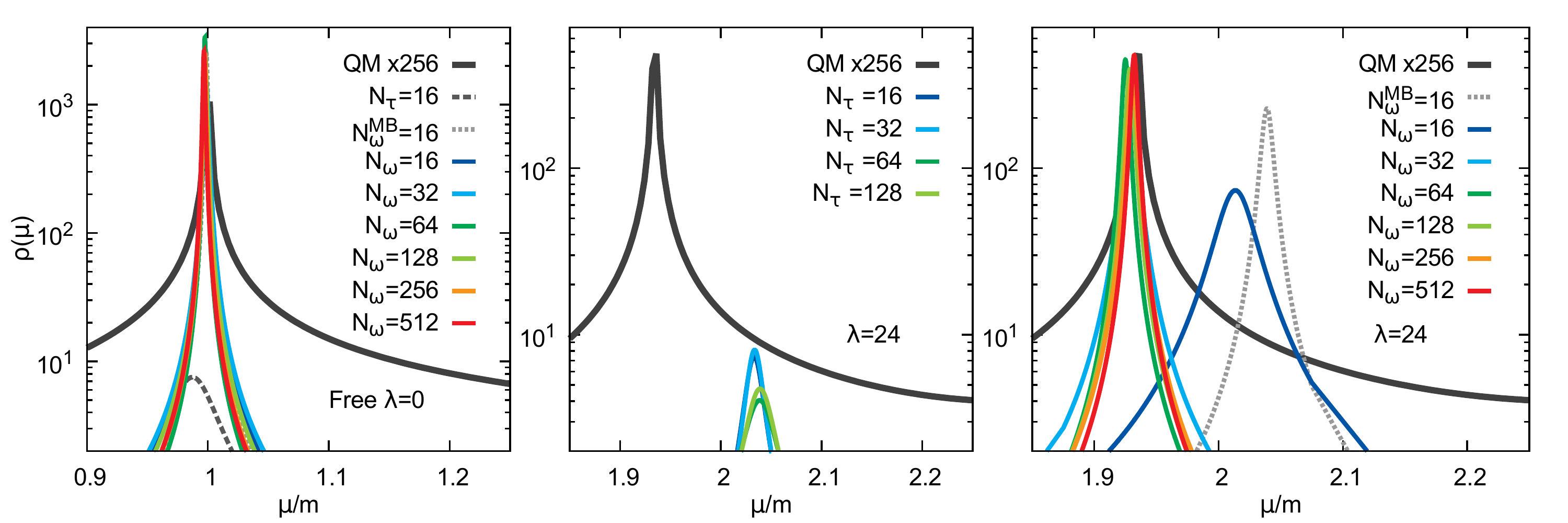}
\caption{Spectral functions of the free theory (left) and at
  $\lambda=24$ unfolded from Euclidean (center) and imaginary
  frequency correlators (right). In each panel $\rho(\mu)$ from the
  corresponding quantum mechanical computation is plotted as solid
  thick gray line. Spectra denoted by $N_\tau$ are reconstructed from
  Euclidean correlators and $N_\omega^{\rm MB}$ from the Fourier
  transformed Matsubara one. The label $N_\omega$ refers to spectra
  from the direct imaginary frequency simulation resolving between
  $\omega_n$.}\label{Fig:SpecRec}
\end{figure*}

In summary, it is the combination of the rational kernel in
\eqref{Eq:FreqRec} and the access to arbitrary frequencies that leads to
exponentially improved accuracy in the spectral reconstruction, see
Fig.~\ref{Fig:SpecRec} for the results.

\begin{figure}[t]
\centering
\includegraphics[scale=0.5, trim= 0cm 0.5cm 0cm 0.4cm,
clip=true]{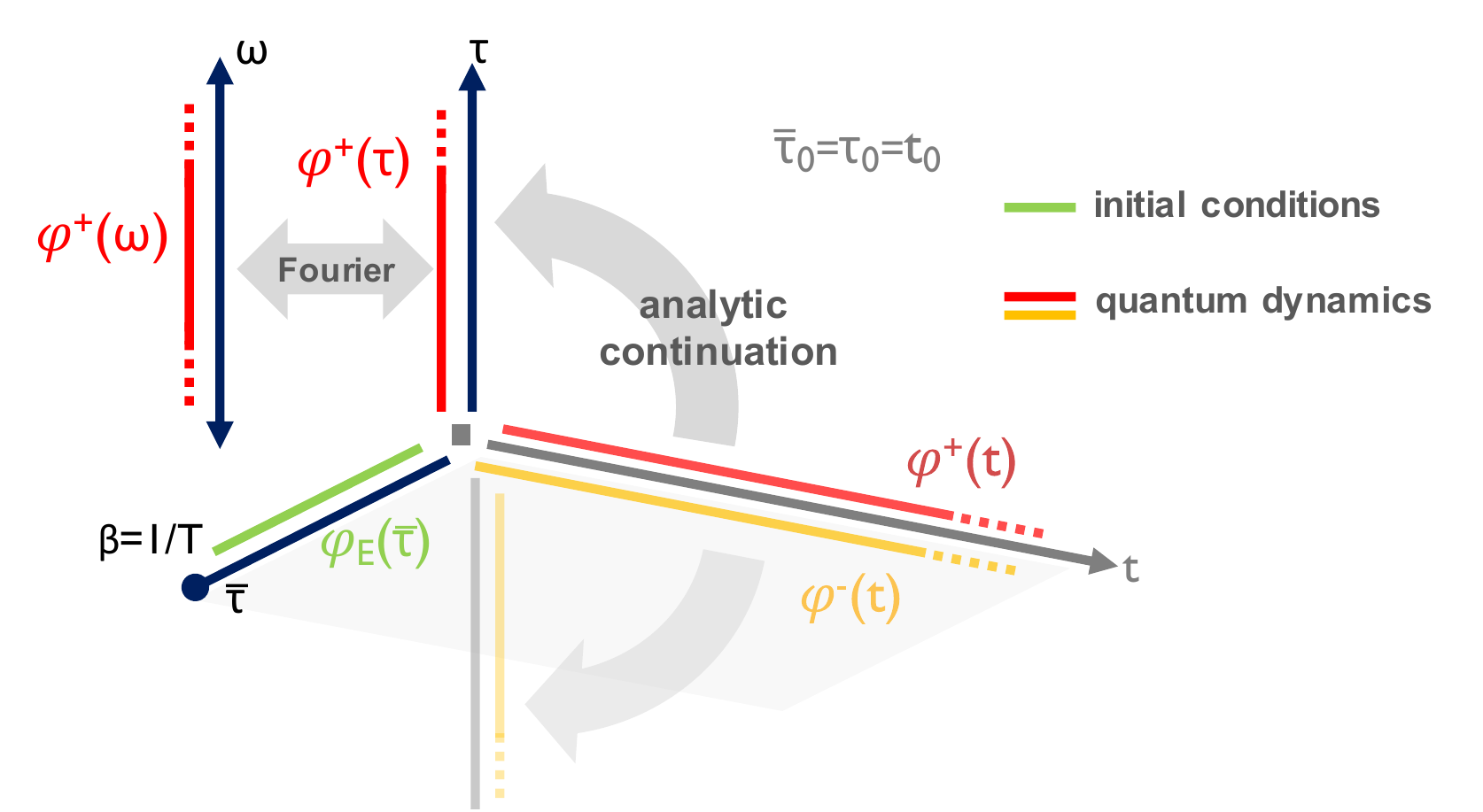}\vspace{-0.2cm}
\caption{Sketch of the imaginary frequency setup.}\label{Fig:Setup}\vspace{-0.4cm}
\end{figure}

Now we discuss the lattice setup that implements the above ideas. The
starting point is a real-time thermal field theory, formulated on the
Schwinger-Keldysh contour \cite{Berges:2006xc,Berges:2015kfa}. It
amounts to an initial value problem with the following path integral
representation for the partition function
\begin{align}
  \nonumber \cal{Z}=&\underbracket{\int_{\varphi_E(0)=\varphi_E(\beta)} {\cal D}\varphi_E  e^{- S_E[\varphi_E]}
                      }_{\rm initial\, conditions}\times \\
                    & \hspace{1.cm}\underbracket{\int_{\varphi^+(t_0,{\bf x})=\varphi_E(0)}^{\varphi^-(t_0,{\bf x})=\varphi_E(\beta)}{\cal D}\varphi
                      e^{iS_M[\varphi^+] - iS_M[\varphi^-]}}_{\rm quantum\, dynamics},
\end{align}
with $S_E$ being the Euclidean- and $S_M$ the Minkowski space action. $\varphi$ lives on the forward and
backward branch, $\varphi^+(t,{\bf x})$ lives on the forward branch
from the inital time $t=t_0=-\infty$ to $t=\infty$, and $\varphi^-(t,{\bf x})$
lives on the backward branch, see Fig.~\ref{Fig:Setup}. The initial conditions are expressed as a
path integral for $\varphi_E(\bar\tau,{\bf x})$ on a compact imaginary
time axis $\bar\tau\in[0,\beta]$. 

Now we use that the real-time contour extends to infinity, and any
correlations between fields at finite t and $t=+\infty$ vanish. This
allows us to cut open the Keldysh contour at $t=+\infty$, and to
rotate the real-time contour to an imaginary time axis, not to be
confused with the time axis of the intial condition path
integral. This rotation is implemented such that the path integrals
for the rotated forward and backward branches have bounded statistical
measures: the forward branch is rotated to the upper complex
half plane while the backward branch to the lower one (see Fig.~\ref{Fig:Setup}).


Furthermore, in equilibrium the correlator $G$ in \eq{Eq:FreqRec} can be derived as
the analytic continuation of the real-time correlator of the fields on the
forward branch, 
\begin{equation}\label{eq:G++--}
G^{++}(\omega):=\langle\varphi^+(\omega)\varphi^+(-\omega)\rangle\,.
\end{equation} 
Thus it suffices to compute the correlation functions on the forward
contour, and we do not consider the backward contour any further
here. The forward branch may now be treated using stochastic
quantization \cite{Namiki:1992wf} with a Euclidean action $S_E$. Its
initial condition is given by the Euclidean field,
$\varphi^+(\tau=t_0)=\varphi_E(0)$, and the correlation function 
$G(\omega)$ can be computed for all $\omega \in \mathbb{R}$. 

\begin{figure}
\includegraphics[scale=0.7]{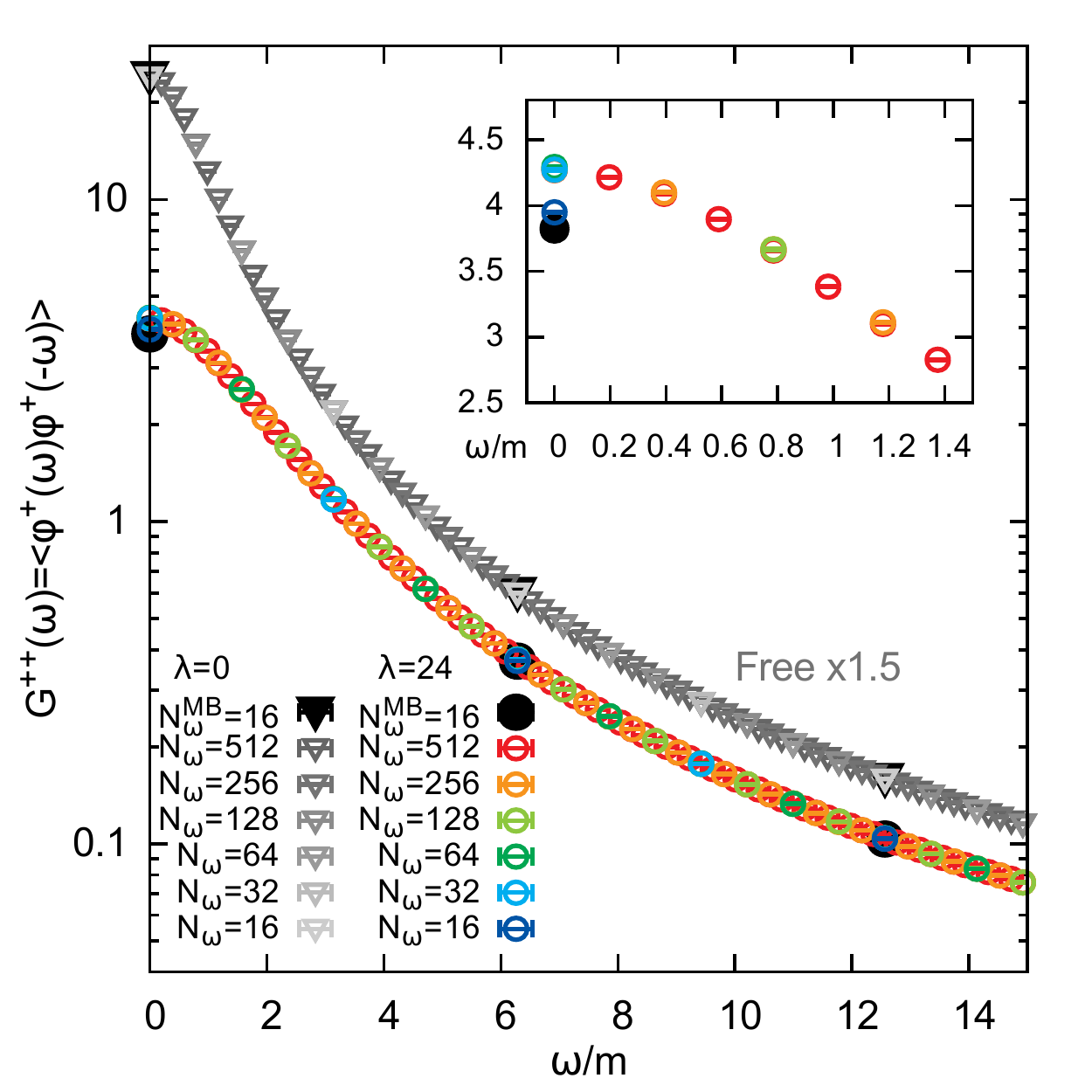}
\includegraphics[scale=0.58]{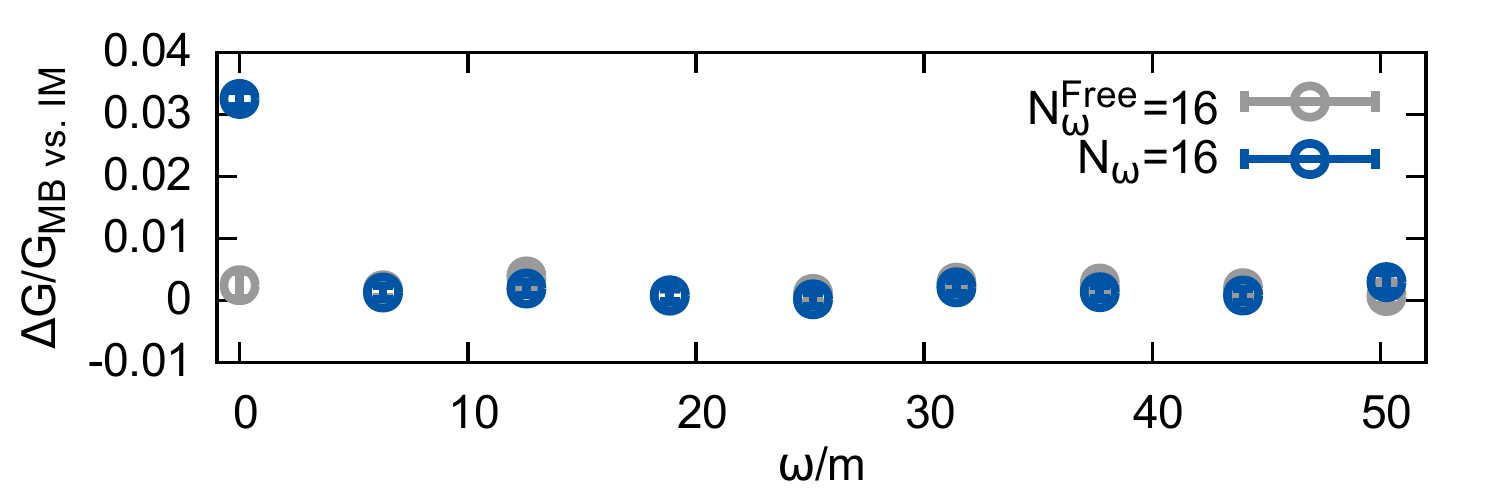}
\caption{ (top) Imaginary frequency correlators $G^{++}(\omega)$ from
  lattice simulations at $\lambda=0$ (gray, shifted $\times1.5$) and
  at $\lambda=24$ (colored). Black filled symbols correspond to the
  correlator at the Matsubara frequencies accessible in standard
  simulations. Open symbols arise from our imaginary frequency
  simulation. The inset shows the low frequency regime at
  $\lambda=24$, in particular that finite volume artifacts are present
  at $\omega_0$ in the Matsubara correlator, which are removed going
  to finer resolution in $\omega$. (bottom) Relative difference
  standard Matsubara correlators and those from our direct
  simulation. Note the excellent agreement except for $\omega_0$ at
  $\lambda=24$, reflecting finite volume artifacts in the standard
  Matsubara value.}\label{Fig:SimFreq}
\end{figure}

In practice the simulation is carried out on a finite temporal lattice with
periodic boundary condition which introduces finite volume effects.
We have checked that these lattice artefacts disappear in the infinite
volume limit, but the convergence is very slow. A qualitatively
enhanced convergence of both, infinite volume and continuum limit, can
be achieved as follows: the Fourier transform to frequency space
translates the initial condition to the constraint
$\sum_l \varphi^+(\omega_l) =\varphi_E(\bar\tau_0)$, where
$|l|=0,...,\lfloor N_\omega/2 \rfloor$, and
$\omega_l= 2 \pi l/(N_\omega \Delta \bar \tau)$. Here
$\Delta \bar \tau$ is the temporal lattice spacing on the standard
Euclidean lattice. This constraint can be rewritten as one for the
field at the highest frequency,
$\varphi^+(\omega_{\lfloor N_\omega/2 \rfloor}) =
\varphi_E(\bar\tau_0)-\sum_l \varphi^+(\omega_l) $. The kinetic term
$1/2\sum_l\varphi^+(\omega_l) {\cal P}(\omega_l)\varphi^+(\omega_l)$
with lattice dispersion $\cal P$ is local in frequency space. Firstly,
the constraint is irrelevant in the continuum limit, as its
statistical weight is vanishing due to
$\varphi^+(\omega_{\lfloor N_\omega/2 \rfloor}){\cal
  P}(\omega_{\lfloor N_\omega/2 \rfloor})\varphi^+(\omega_{\lfloor
  N_\omega/2 \rfloor})\to \infty$. Secondly, in the infinite volume we
have an infinite number of momentum modes. This entails that the
weight of one specific momentum mode tends towards zero.  In summary,
the dependence of the kinetic term on the inital condition is a
well-understood lattice artefact. In the spirit of improved lattice
actions we can lift the Matsubara constraint for all terms in the
lattice action that are local in frequency space. Here we apply this
reasoning to $S_0$, the quadratic part of the action. The efficient
convergence of this approach is
confirmed for the propagator, see Fig.~\ref{Fig:SimFreq}.

The full theory is then formulated in a mixed representation in
$\tau$ and $\omega$. While $S_0$ is simulated in $\omega$-space,
$S_{\rm int}=S-S_0$ is directly implemented via a discrete Fourier
transform in $\tau$-space, since interactions are local on the
sites. Approximating the convolutions in the interaction term via DFT
introduces a finite volume artifact, which vanishes as
$N_\omega\to\infty$. The thermal initial condition
$\varphi^+(\tau_0)=\varphi_E(\bar\tau_0)$ solely enters via this
interaction term. This concludes the formulation and discussion of our
lattice setup.

Our novel approach is now tested in a $0+1$ dimensional scalar
field theory at temperature $T=1/\beta$, whose Euclidean action reads
\begin{equation}
  \nonumber S_E= \int d\tau \big( \underbracket{
    \frac{1}{2}(\partial_\tau \varphi_E)^2 + \frac{1}{2}m^2
    \varphi_E^2}_{S_E^0} +\underbracket{\frac{\lambda}{4!}
    \varphi_E^4}_{S_E^{\rm int}} \big)\,.
\end{equation}\vspace{-0.3cm}

The system is simulated on a frequency lattice with $N_\omega$ points
using stochastic quantization in Langevin time s,
\begin{equation}\label{eq:SQpgi+}
  \partial_{s} \varphi^+(\omega_l)=-\frac{\delta S_0}{\delta
    \varphi^+(\omega_l)} - \frac{\delta S_E^{\rm int}}{\delta
    \varphi^+(\tau_j)}\frac{\delta
    \varphi^+(\tau_j)}{\delta
    \varphi^+(\omega_l)}+\eta(\omega_l)\,. 
\end{equation} 
Here a sum over $j=1,...,N_\omega$ is implied and the dispersion reads
${\cal P}(\omega_l)=(2-2{\rm cos}[2\pi l /N_\omega ])/\Delta \bar
\tau^2$, $l\in [0,N_\omega-1]$. In \eqref{eq:SQpgi+} $\eta(\omega_l)$
is the Fourier transform of the Gau\ss ian noise, $\langle
\tilde\eta(\tau_i)\tilde\eta(\tau_j)\rangle=\,2\delta_{ij}/
\Delta\bar\tau$ and $S_0$ represents the discretized free action,
\begin{equation}\label{eq:S0}
  S_0=\sum_l \Delta\omega\; \frac{1}{2}\big({\cal P}(\omega_l) + m^2\big)
  |\varphi^+(\omega_l)|^2\,,  
\end{equation}
while $S_E^{\rm int}=S_E-S_0$ is the interaction term.  Concurrently
we sample the thermal initial conditions at $N_\tau$ points along the
compactified Euclidean time $\tau$,
\begin{equation}
\partial_{s} \varphi_E(\bar\tau_k)=\frac{- \delta
    S_E}{\delta \varphi_E(\bar\tau_k)} + \eta(\bar\tau_k) ,\; \langle
  \eta(\bar\tau_i),\eta(\bar\tau_j) \rangle =
  2\frac{\delta_{ij}}{\Delta\bar\tau}\,.\label{Eq:StochQuant}
\end{equation}
so that the value of $\varphi^+(\tau=\tau_0)$ can be replaced with
$\varphi_E(\bar\tau=\bar\tau_0)$ each time when computing
$\frac{\delta S_E^{\rm int}}{\delta \varphi^+(\tau_j)}$ in
\eqref{eq:SQpgi+}. After $\Delta s=1$ we record the correlators
$G_E(\bar\tau)=\langle \varphi_E(\bar\tau)\varphi_E(0)\rangle $
and $G^{++}(\omega)=\langle \varphi^+(\omega)\varphi^+(-\omega)\rangle
$, collecting in total $N_{\rm conf}=10^6$ samples, which leads to a
relative statistical error $\Delta G/G\sim10^{-3}$.

To crosscheck our computations we have also performed an equivalent quantum
mechanical (QM) computation of the thermal real-time correlation
functions $G^{\rm QM}(t)\equiv \langle \hat{x}(t)\hat{x}(0)\rangle$
using the anharmonic oscillator in the truncated Hilbert space of
$32$ energy eigenfunctions of the harmonic oscillator
\cite{Berges:2006xc}. The spectral function is obtained from a Fourier transform of
${\rm Im}[\langle \hat{x}(t)\hat{x}(0)\rangle]$ along
$N_t=3.2\times10^4$ steps of $\Delta t=0.05$, i.e.\ it is computed only
over a finite real-time extent. Due to the restricted Hilbert space
and integration this computation underestimates the height of
delta-peak like structures, present e.g. in the free spectrum and
contains ringing.

\begin{figure}
  \hspace{0.2cm}\includegraphics[scale=0.57, trim=0 0 0 0,
  clip=true]{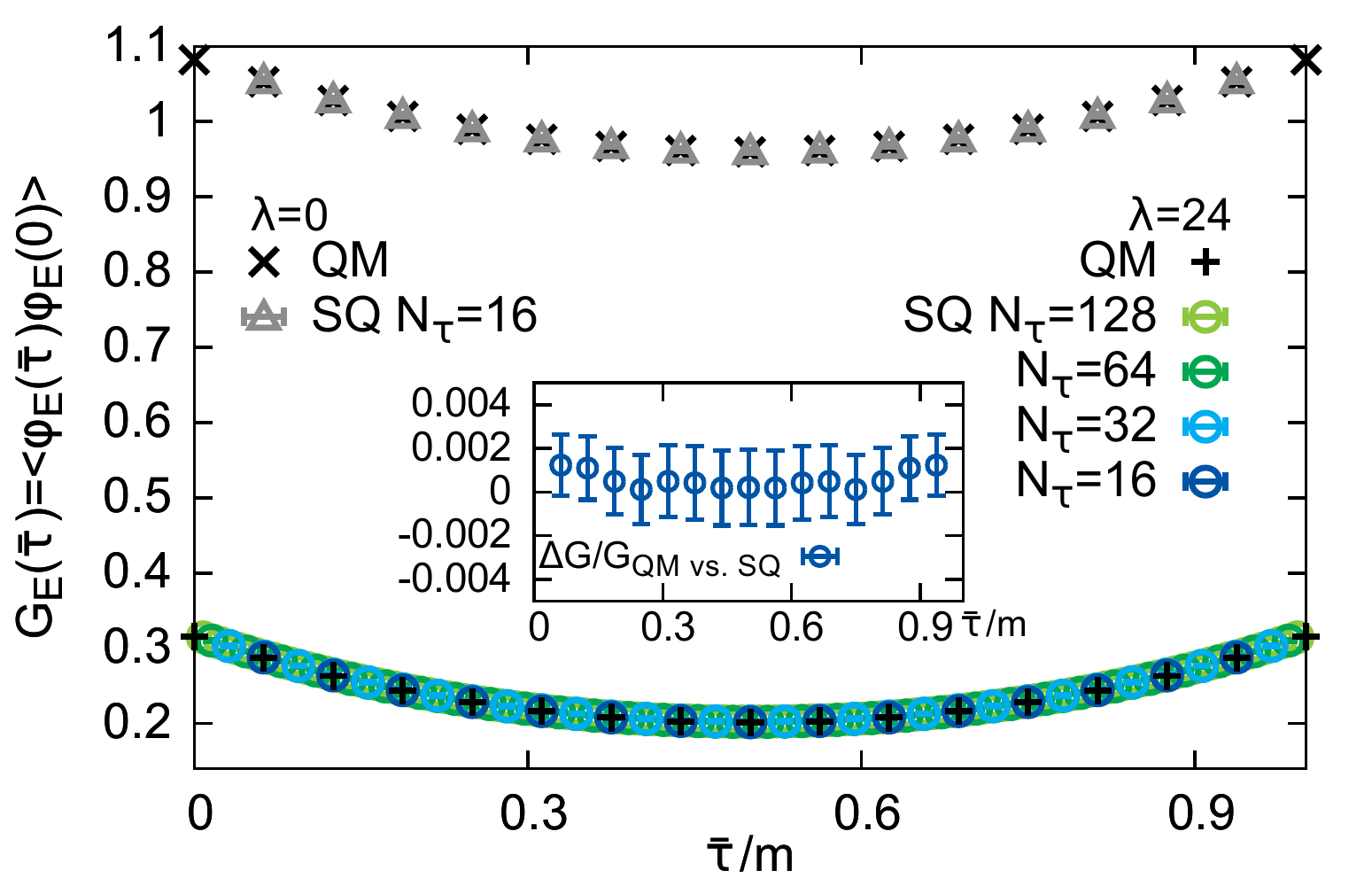}
\caption{Euclidean correlation functions $G_E(\bar\tau)$ for $\lambda=0$
  (upper points) and $\lambda=24$ (lower points). Black crosses denote
  correlators from a quantum mechanical computation, while open
  symbols represent standard lattice simulations. The inset shows the
  relative difference between the lattice result for $\lambda=24$ at
  $N_{\bar\tau}=16$ and the quantum mechanical
  computation.}\label{Fig:SimEucl}
\end{figure}

In Fig.~\ref{Fig:SimEucl} we show the Euclidean correlators computed
for the thermal scalar system at $\beta=1$ and $m=1$ along the
standard compactified Euclidean axis for the free $\lambda=0$
(top points) and the interacting case $\lambda=24$ (bottom
points). The black crosses denote the outcome of the QM computation and open symbols denote lattice results. For
the free case we only carry out simulations at $N_{\bar\tau}=16$, while at
$\lambda=24$ we produce data at four different
$N_{\bar\tau}=16\ldots128$. The inset shows the relative difference between
the interacting correlator at $N_{\bar\tau}=16$ computed from the lattice
and via quantum mechanics. For our choice of $ds=10^{-4}$ they agree
within the statistical errors of the lattice simulation. Note that increasing
$N_{\bar\tau}$, i.e. a smaller $\Delta \bar{\tau}$, leads to a larger drift term, so that
at $N_{\bar\tau}=128$ the $ds$ needs to be reduced by two orders of magnitude
to maintain an accurate outcome.


We focus on simulations with $N_{\bar\tau}=16$, and inspect the
corresponding correlation functions in imaginary frequency space given
in the top panel of Fig.~\ref{Fig:SimFreq} for both the free (gray) and
interacting (colored) case. We show only the relevant low frequency
regime up to $\omega/m=15$ while the simulations reach up to
$\omega_{8}/m\approx50$. As $\varphi^+(\tau)\in\mathbb{R}$ the
correlators are real and symmetric around $\omega=0$.

The black filled symbols denote the correlators obtained from simply
Fourier transforming the fields along the compact Euclidean domain and
thus are located at the Matsubara frequencies $\omega_n$. The open
symbols are obtained from the fields simulated directly along the
imaginary frequency domain using between $N_\omega=16\ldots512$
points. In the free $\lambda=0$ case our novel simulation yields a
smooth interpolation between the Matsubara frequencies and furthermore
as can be seen from the gray points in the lower panel of
Fig.~\ref{Fig:SimFreq} agrees excellently with the values obtained from
the Fourier transform of the compact Euclidean domain. Note that at $\lambda=0$ the
two concurrent simulations \eqref{eq:SQpgi+} and \eqref{Eq:StochQuant}
are completely decoupled.

In the interacting case denoted by colored open symbols in
Fig.~\ref{Fig:SimFreq}, we again find a smooth interpolation between
Matsubara frequencies and very good agreement with the standard
Matsubara propagators for $\omega_{n>0}$. Let us inspect the regime
close to the origin, given by the inset on the top of
Fig.~\ref{Fig:SimFreq}. Already at $N_\omega=16$ the new simulation
(dark blue) leads to a value at $\omega=0$ that differs from the
conventional Matsubara one (black). Increasing $N_\omega$ quickly lets
the correlator converge significantly above the standard
estimate. Since in a lattice simulation $\omega=0$ reflects the sum
over all corresponding imaginary time extent, we infer that the
standard simulation along the compactified Euclidean domain is not
able to capture all the relevant late Euclidean-time physics that
however can be vital to thermal processes. Its purpose is only to
correctly sample $\varphi(\bar\tau_0)$, which it does. 

Now we extract spectral functions from the correlators, with kernels
$\tilde{K}(\mu,\bar\tau,T)={\rm cosh}[\mu(\bar\tau-/2T)]/{\rm
  sinh}[\mu/2T]$ and $K(\mu,\omega)=\frac{2\mu}{\mu^2+{\cal
    P}(\omega)}$. The latter is the lattice analogue of the continuum
K\"all\'en-Lehmann kernel, and its use leads to a more efficient
convergence of the spectral reconstruction. Here we deploy a recent
Bayesian reconstruction (BR) method \cite{Burnier:2013nla} and
discretize positive real-time frequencies $\mu\in[10^{-3},200]$ with
$N_\mu=4000$ points including a high frequency window of $N_\mu^{\rm
  HP}=1000$ around the lowest lying peaks. For minimal bias we use a
flat default model $m(\mu)=1$. In the free case we know that only a
single delta-peak at the mass of the field exists in the spectrum. The
quantum mechanical computation given by the thick gray line in the
left panel of Fig.~\ref{Fig:SpecRec} while being able to capture the
position of the peak, gives an artificial finite width due to the
underlying finite real-time extent Fourier transform. We have checked
that it can be reduced systematically and therefore for visualization
purposes multiply the curve here by a factor of 256. The
reconstruction from the $N_{\bar\tau}=16$ compact Euclidean domain
(dark gray dashed) both gives a too large width and its position is
still slightly off. A Fourier transform of the Euclidean fields and
reconstructing along Matsubara frequencies already improves the
results significantly. The light gray dashed line shows a much reduced
width and is positioned very close to unity. Since here the standard
Matsubara data already leads to a very good result, the improvement
from using the correlators of the new simulation only leads to a
further decrease in the reconstructed width. Note that the width
depends on the error of the data and can be improved by more
statistics.

Turning to the interacting case $\lambda=24$, we first carry out
the reconstruction from Eucldiean data, using different resolutions
along $\bar\tau$. In the center panel of Fig.~\ref{Fig:SpecRec} the
results are shown for $N_{\bar\tau}=16\ldots128$. We find that the
reconstruction does not capture the position of the lowest lying peak
in the spectrum and even more importantly does not show a systematic
improvement with $N_{\bar\tau}$. The reason is that with larger $N_{\bar\tau}$
higher $\omega_n$ become accessible but contribute only marginally to
the relevant physics.

If on the other hand we use data in imaginary frequency space the
situation is very different as shown in the right panel of
Fig.~\ref{Fig:SpecRec}. Again simply using the standard Matsubara
correlators already improves the width of the reconstruction while the
peak position is still off (gray dashed). However the access to the
correlator at frequencies between $\omega_0$ and $\omega_1$ allows us
to systematically and significantly improve the reconstruction of the
lowest lying peak, with $N_\omega=512$ reproducing the peak position
with high accuracy.

In summary we have shown that by carefully distinguishing initial
conditions from quantum dynamics it is possible to devise a simulation
prescription for thermal quantum fields along general imaginary
frequencies. In particular this approach gives access to the thermal
physics between $\omega_0$ and $\omega_1$. Real-time spectral
functions are related to imaginary time correlators via an integral
transform with a rational kernel, and in combination their extraction
becomes exponentially improved. The success of this approach has been
demonstrated here with scalar fields. It can be straightforwardly
extended to Abelian and non-Abelian gauge theories which is work in
progress. Evidently, computations of phenomenologically relevant
quantities such as transport coefficients and heavy quark spectral
functions will be improved significantly with our simulation
prescription.

We thank N.~Christiansen and S. Fl\"orchinger for discussions. A.R.\
acknowledges fruitful exchanges with P.~de Forcrand and A.~Kurkela at
CERN and during the ECT* workshop on advances in transport. This work
is supported by EMMI, the grants ERC-AdG-290623, BMBF 05P12VHCTG.  It
is part of and supported by the DFG Collaborative Research Centre "SFB
1225 (ISOQUANT)".

\bibliographystyle{bibstyle}
\bibliography{../bib_lat-spec}

\end{document}